# Long-haul coherent communications using microresonator-based frequency combs

ATTILA FÜLÖP,[1,*] MIKAEL MAZUR,[1] ABEL LORENCES-RIESGO,[1] TOBIAS A. ERIKSSON,[1,†] PEI-HSUN WANG,[2] YI XUAN,[2,3] DAN E. LEAIRD,[2] MINGHAO QI,[2,3] PETER A. ANDREKSON,[1] ANDREW M. WEINER,[2,3] AND VICTOR TORRES-COMPANY[1]

[1]Photonics Laboratory, Department of Microtechnology and Nanoscience, Chalmers University of Technology, SE-41296 Göteborg, Sweden
[2]School of Electrical and Computer Engineering, Purdue University, West Lafayette, IN 47907-2035, USA
[3]Birck Nanotechnology Center, Purdue University, West Lafayette, IN 47907-2035, USA
[†]Now at Nokia Bell Labs, Stuttgart, Germany

*Corresponding author: attila.fulop@chalmers.se

**Microresonator-based frequency combs are strong contenders as light sources for wavelength-division multiplexing (WDM). Recent demonstrations have shown the potential of microresonator combs for replacing tens of WDM lasers with a single laser-pumped device. These experiments relied on microresonators displaying anomalous dispersion. Devices operating in the normal dispersion offer the prospect of attaining high power conversion efficiency - an aspect that will be crucial in the future for enabling energy-efficient coherent communications with higher order modulation formats or lighting several spatial channels in space-division multiplexing. Here we report the experimental demonstration of coherent communications using normal dispersion microresonator combs. With polarization multiplexed (PM) quadrature phase-shift keying, we transmitted data over more than 6300 km in single-mode fiber. In a second experiment, we reached beyond 700 km with PM 16 quadrature amplitude modulation format and an aggregate data rate above 900 Gbit/s assuming 6% error correction overhead. These results represent the longest fiber transmission ever achieved using an integrated comb source.**

The advent of self-referenced femtosecond mode-locked lasers [1, 2] allowed for establishing a coherent link between radio frequency and optical signals obtained from lasers or atomic transitions. This technology is finding an increasing number of applications in precision frequency synthesis and metrology [3]. Microresonator frequency combs have been identified as a key technology to reach a similar level of precision in a monolithic platform. This technology provides an opportunity to attain line spacings significantly higher than what can be achieved with standard mode-locked lasers [4, 5]. In addition, high-Q microresonators can be fabricated using standard semiconductor fabrication processes, see e.g. [6-11]. As a result, microresonator frequency combs are opening up a whole new range of technological possibilities. Recent demonstrations include self-referencing [12, 13], optical clocks [14], radio-frequency photonics [15, 16], spectroscopy [17, 18], optical waveform synthesis [19] and high-capacity communications [20].

In fiber-optic communications, an integrated multi-wavelength light source would provide the fundamental carriers on which to encode data via e.g. wavelength-division multiplexing (WDM), see Fig. 1. A frequency comb would allow replacing tens of WDM channel light sources with a single laser [21-23]. Absolute frequency accuracy is in principle not needed, but line spacing stability provides an opportunity to mitigate inter-channel nonlinear distortion via digital back propagation [24], a prospect not achievable using a laser array. Since a self-referenced light source is not necessary, there is a multitude of integrated comb sources that can be considered, such as integrated electro-optic combs [25], silicon-organic hybrid modulators [26], passively mode-locked lasers [27] or quantum-dash mode-locked lasers [28]. The rationale for using microresonator-based frequency combs lies in the fact that they can be generated using high-performance silicon nitride high-Q microresonators [8, 11, 29]. This technology is compatible with standard CMOS fabrication processes and recent demonstrations have shown that it is possible to realize multi-layer integration with active silicon components [30]. This significant achievement opens up the possibility to in the future realize a fully integrated comb-based transceiver with silicon photonics technology. Indeed, silicon nitride microresonator combs have shown a level of performance compatible with coherent communications [20]. Bright temporal solitons in silicon nitride microresonator combs have led to impressive demonstrations attaining 50 Tb/s aggregate data rates [31]. These results demonstrate the potential to achieve hundreds of lines in a single device by proper dispersion engineering.

A fundamental challenge however with bright temporal solitons is the power conversion efficiency [32], typically below 1% [33, 34]. This limits the amount of power per line that can be attained. As a result, the power conversion efficiency provides a fundamental limit in the signal-to-noise ratio per WDM channel given a fixed pump power. This could result e.g. in a limited reach using complex modulation formats. Recent theoretical [35] and experimental investigations [36] point out that microresonator cavities designed to operate in the normal dispersion regime can lead to combs with substantially higher power conversion efficiencies, e.g. >30% as demonstrated in [36] The performance of normal dispersion combs has however not been assessed in the context of coherent optical communications.

In this Letter we demonstrate the results from two long-haul communication experiments using microresonator combs designed in silicon nitride waveguides exhibiting normal dispersion. We modulated data using polarization-multiplexed quadrature-phase shift keying (PM-QPSK) and 16-quadrature amplitude modulation (PM-16QAM) on two different comb devices. We achieved a propagation distance of 6300 km using 12.5 GBaud PM-QPSK and 700 km with 20 GBaud PM-16QAM. This represents the longest transmission distance ever achieved using an integrated comb source.

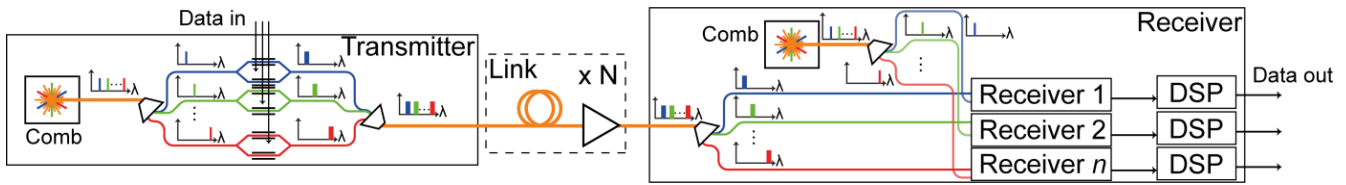

Fig. 1. In an ideal comb-based coherent optical communications link, both the transmitter and the receiver contains a frequency comb to be used as the initial light source as well as the local oscillator respectively. After separately modulating the comb lines in the transmitter using $n$ separate modulators, they would be recombined and transmitted down a link consisting of $N$ fiber spans and amplifiers. After the link they would again be separated, received and decoded separately using the receiver comb as local oscillator.

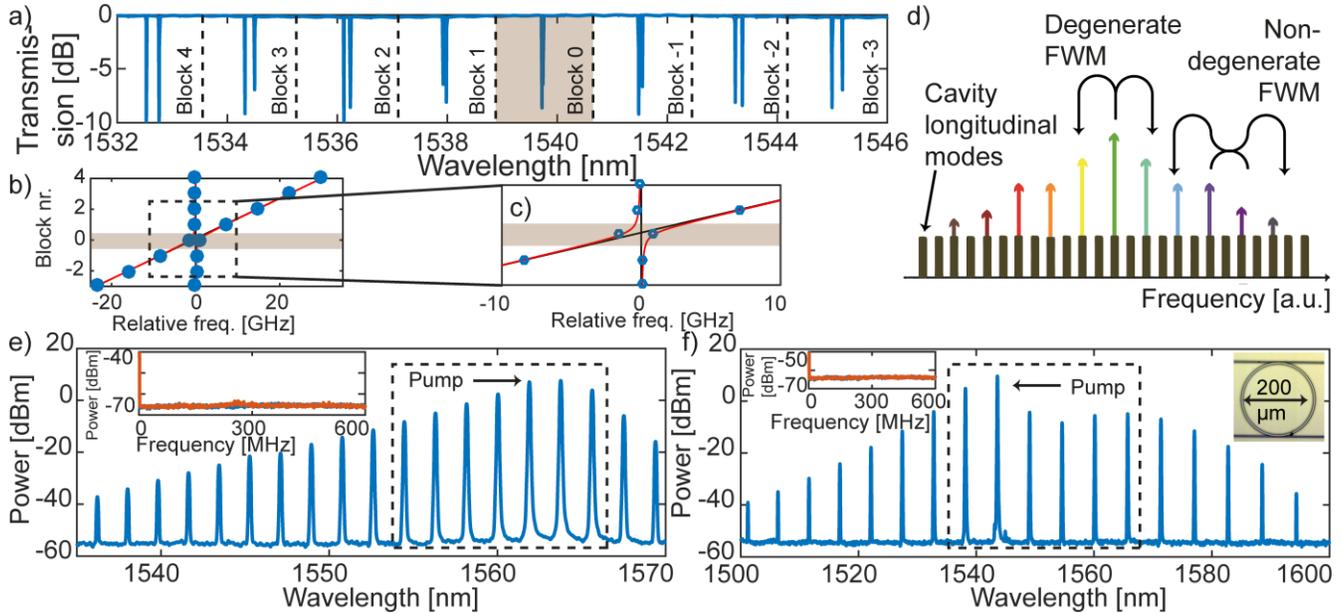

Fig. 2. a) A transmission scan of a multimode device highlighting a region where two modes couple linearly. By dividing the spectrum into regions spaced with 1 FSR, we can illustrate the coupling effect in subfigures b) and c). As the resonance locations move closer, they will repel each other if there is coupling between them, leading to the resonances being slightly offset in the shaded region. Since this is locally changing the effective index of the waveguide, the dispersion of the device will also be greatly affected. A more detailed description of the effects of modal coupling is given in [38]. d) Locally anomalous-dispersion enables degenerate four-wave mixing to initialize the comb generation. After the initial three lines are present, non-degenerate four-wave mixing produces additional lines. e) and f) The optical power spectrum of the two used combs with 230 GHz and 690 GHz line spacing respectively as seen from their drop ports with the dashed boxes showing the lines used in the communications experiments. The left insets display the corresponding RF spectrum in red with the noise floor in blue showing that the combs were operating in a low-noise state. The right inset in f) shows a microscope image of a microresonator.

The devices consisted of rings with 100 μm radius and were equipped with a drop port, as shown in Fig. 2g). The drop port assisted in filtering out amplified spontaneous emission (ASE) noise that was left around the pump laser while at the same time limiting the amount of pump light that was present at the output port [37]. Contrary to the cavity soliton-based microresonator combs that require anomalous dispersion waveguides, these microresonators were designed to operate in the normal dispersion regime, permitting slightly thinner waveguides at 600 nm thickness. To initialize the combs from a single continuous wave (CW) laser source using degenerate four-wave mixing, it is however still required to have locally anomalous dispersion. This can be achieved through local perturbations. In our case, the perturbation was caused by coupling between transverse modes [38]. Figures 2b) and 2c) illustrate the effect of modal coupling where in certain frequency regions the resonance locations, and thus the effective index, is strongly perturbed by the presence of a second (higher order) mode. This leads to a locally anomalous dispersion, enabling the degenerate four-wave mixing process in Fig. 2d). Figure 2e) and 2f) show the optical spectra of the combs that were used in our experiments. The RF spectra in the insets indicate that the combs were operating in a low-noise state [39]. The total conversion efficiencies (adding the through and the drop port) of both used combs were measured to be above 10% (see Supplementary materials section 2).

Before modulating data onto the comb lines, we also placed a flattening stage consisting of a programmable pulse shaper and an EDFA ensuring that all lines had equal power. This resulted in a flat spectrum with an OSNR of >35 dB per line at 0.1 nm resolution before data modulation. To emulate the system in Fig. 1 in the laboratory without the large amount of components required, all the lines of the frequency comb were instead modulated simultaneously using the same modulator. A recirculating loop with two 80 km fiber spans inside was used as the transmission link. Further, at the receiver end, we decoded one channel at a time using a free-running single laser as local oscillator and a tunable optical filter to filter out the matching channel. A simplified sketch of the setup is shown in Fig. 3a) while the full implementation details of the recirculating loop and the transmission setup are described in Supplementary materials section 1. By using a recirculating loop we also had the ability to vary the transmission distance, allowing us to find the maximum permitted distance given the chosen pre-forward error correction (FEC) bit error rate (BER) $10^{-3}$.

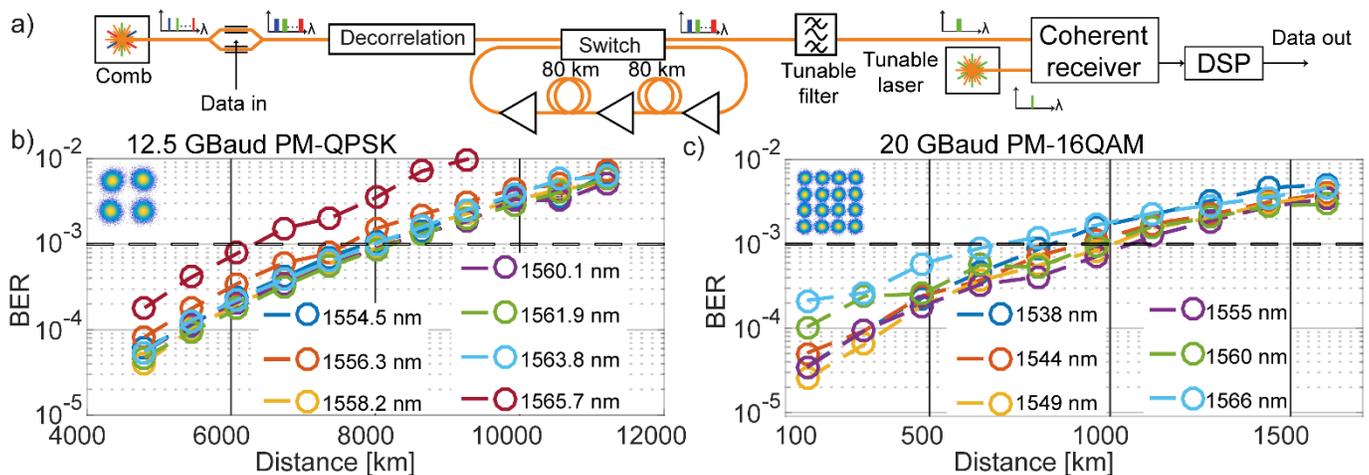

Fig. 3. a) Actual lab implementation of the long-haul link consisting of a frequency comb source connected to a modulator, a decorrelation stage, a recirculating loop and a coherent receiver. b) and c) Result plots of the PM-QPSK and the PM-16QAM transmission experiments showing BER as a function of distance in the recirculating loop. The insets display received and decoded signal constellations with bit error rates below $10^{-3}$.

The two different modulation formats were tested with two slightly different transmitters due to lab constraints. In the QPSK case, a single free spectral range (FSR) spaced comb (corresponding to 230 GHz line spacing) with 7 lines at a part of the C-band (see the dashed box in Fig. 2e)) was modulated at 12.5 GBaud using a pseudo-random bit sequence (PRBS) pattern generator. In the 16QAM system a second microresonator giving a 3 FSR spaced comb (corresponding to 690 GHz line spacing), covering the full C-band with 6 lines (see Fig. 2f)) was modulated with the use of PRBS and 16QAM at 20 GBaud using an arbitrary waveform generator. Both setups used the same recirculating loop as well as receiver components, described in more detail in Supplementary materials section 1.

The two chosen modulation formats and symbol rates permitted transmission over different distances. With the QPSK system, at a cumulative pre-FEC bit rate of 350 Gbit/s, we reached more than 6300 km with all the lines, three of which reached beyond 8000 km. Since the higher bit rate 16QAM system (at a cumulative pre-FEC bit rate of 960 Gbit/s) requires a higher OSNR at the receiver [40], the permitted distance for the second experiment was instead 700 km with three lines going beyond 950 km. Figures 3b) and 3c) show how the BER varied as a function of transmitted distance for both systems. In both cases, the outermost line at 1566 nm suffered a penalty in terms of distance. To ensure that this was not due to excess noise in the comb line itself, we performed back-to-back noise loading measurements. The measured implementation penalty of the comb lines did not differ significantly from that of free-running lasers, see Supplementary materials section 1.

In summary, we have demonstrated experimentally that long-haul transmission is possible using silicon nitride microresonator-based frequency comb technology. Owing to the conversion efficiency enabled by normal dispersion combs, we managed to transmit data over record distances. The low-complexity QPSK format permitted transmission over transatlantic distances. The higher symbol rate 16QAM also permitted long-haul communications albeit at shorter distances. In the future, more WDM channels will be needed to fill the whole C band with a line spacing compatible with the ITU grid. We envision that the use of mode-locked dark pulses in normal dispersion combs [41], with recently demonstrated conversion efficiencies >30% [36] will provide a viable solution to achieve a favorable scaling with the number of channels and attain high spectral efficiency. The results of this Letter widen the scope of the usability of silicon photonics-based combs from short-reach high-capacity links to long-haul links, thus opening up the path to chip-based transceivers where the coherent nature of frequency combs can be taken advantage of.

**Funding.**

The European Research Council (ERC-2011-AdG-291618 PSOPA); the Swedish Research Council (VR); the KA Wallenberg foundation; the National Science Foundation (NSF) (ECCS-150957); DARPA (W31P40-13-1-001.8); and AFOSR (FA9550-15-1-0211).

See Supplement 1 for supporting content.

# Supplementary information

## 1. LONG-HAUL TRANSMISSION SETUP

### A. The comb generation

The combs were pumped using tunable external cavity lasers (specified linewidth below 100 kHz) amplified using a high-power erbium-doped fiber amplifier (EDFA). To avoid excess amplified spontaneous emission noise, a 1 nm optical bandpass filter was placed before the microresonator. The comb used for the QPSK experiment was pumped by 27 dBm (corresponding to $\sim$ 24 dBm on-chip power) while the comb used for the 16QAM experiment was pumped with 30 dBm (corresponding to $\sim$ 27 dBm on-chip power). The comb operation was stabilized by using a temperature-controlled stage (limiting the temperature variations to less than 0.01 °C) as well as lensed fibers that were pushed into U-grooves on the chip [1]. Using the grooves ensured that the fiber-to-chip coupling remained stable over many hours. Following the chip, a flattening stage consisting of an optical pulse shaper and a second EDFA ensured that all the used lines had equal power going into the modulator. Figure S1a) shows a schematic of both the comb generation and the transmitter.

### B. The transmitter

The transmitter consisted of an electro-optic IQ-modulator modulating all the comb lines simultaneously. The modulator was followed by a dual-polarization emulation stage using a polarization maintaining fiber (PMF) corresponding to a delay of more than 20 symbols to emulate dual-polarized signals. The intrinsic dispersion of a 27 km single-mode fiber was used to delay and decorrelate the different channels with respect to each other. Two different transmitters were connected to the modulator in the two experiments. For the longest reach one, a 12.5 GBaud QPSK signal was generated using a compact pseudorandom bit sequence (PRBS) pattern generator. For the higher bit rate measurement, a more spectrally efficient QAM signal was used. This 20 GBaud 16QAM signal was generated using an arbitrary waveform generator.

Both comb and transmitter combinations were tested in a back-to-back noise loading setup, where progressively more ASE noise was added to the signal after modulation. The optical signal-to-noise ratio (OSNR) at which a bit error rate (BER) of $10^{-3}$ was received was then recorded for each comb line. Reference measurements were done where free-running lasers were instead tuned to the wavelength locations matching the comb lines. This was done to filter out any implementation penalty caused by the combs themselves. Figure S1b) shows the results for the two systems. The conclusion is that the comb lines do not cause any significant penalty. The permitted transmission distance will therefore be limited by the comb lines' initial OSNR and the imperfections in the various other pieces of equipment that are part of the setup.

### C. The recirculating loop

The long-haul fiber link was emulated by using a recirculating fiber loop. Figure S2a) shows a schematic of the recirculating loop and its components. The loop contained two spans of 80 km single-mode fiber (SMF). The fiber spans were preceded by a flattening stage and a tunable attenuator ensuring that the launch power of each line in the comb was kept at optimum. For the QPSK case the optimum total launch power was found to be 4 dBm while for the shorter distance 16QAM experiment it was 6.5 dBm. The pulse shapers used for flattening also attenuated any out-of-band noise. Each SMF span was followed by an EDFA to compensate for the losses. Additionally, the loop also contained a polarization scrambler synchronized to the switches and the loop roundtrip time. This way, by iterative optimization of the flattening stages it was possible to transmit the full combs over multiple tens of roundtrips.

### D. The receiver

At the end, there was a standard coherent receiver stage. One comb line was filtered out at a time and combined with a tunable local oscillator. A dual-polarization optical hybrid followed by a coherent receiver was connected to four ports of a 50 GSamples/s, 23 GHz bandwidth real-time oscilloscope, see Fig. S2b). Batches of 2 million samples were recorded and processed offline. First the receiver imbalance was compensated for, after which the data was resampled to twice the baud rate. The dispersion from the fiber link was then compensated for after which an adaptive equalizer and a frequency and phase estimator performed the final processing. For the QPSK signal, a constant modulus algorithm was used together with an FFT-based frequency offset estimator and a Viterbi-Viterbi phase estimator. In the 16QAM case, a decision-directed least mean square equalizer was used with an FFT-based frequency offset estimator and a blind phase search algorithm [2]. Finally the BER was calculated by comparing the decoded data with the known transmitted sequences.

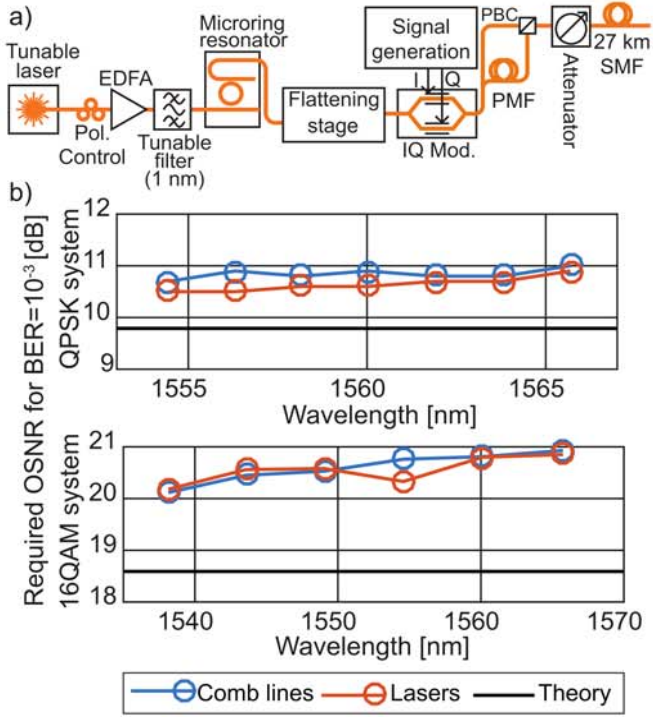
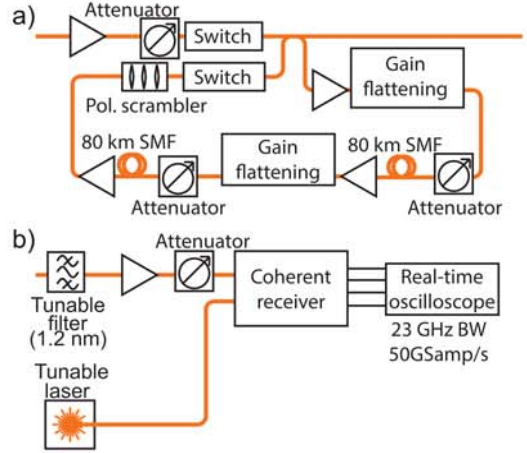

**Fig. S1.** a) Schematic of the equipment used for the comb generation and the data modulation. The PMF delay and polarization beam combiner (PBC) were used to emulate a dual polarization transmitter. b) Results for back-to-back noise loading measurements of both systems, the QPSK transmitter had an implementation penalty of $1.0 \pm 0.2\,\text{dB}$ while the 16QAM transmitter had an implementation penalty of $1.9 \pm 0.4\,\text{dB}$.

## 2. COMB CONVERSION EFFICIENCY

The most straight-forward way of defining conversion efficiency is the relation between the optical power in all the newly generated comb lines (excluding the remaining pump line power) compared to the power in the initial pump [3]. Both measurements should be done in the bus waveguide, adjacent to the ring:

$$\eta_{\text{comb}} = \frac{\sum P_{\text{comblines}}}{P_{\text{pump}}}. \quad \text{(S1)}$$

Since the optical powers inside the microresonator device cannot be measured directly, we have to extract the relevant information from measurements done at the output ports of our device. Figure S3a) shows a schematic of the device, the locations where the optical spectrum should be measured and the different loss elements that are not part of the comb generation process itself. Since the device is coupled into using tapered fibers at both ends, we will incur coupling losses at both facets, $\alpha_{c1}$ and $\alpha_{c2}$. In addition there will also be some (possibly insignificant) losses due to absorption in the bus waveguide, $\alpha_{\text{wg}}$. In practice, the spectrum can only be measured at the input and the output. The total loss can be measured by pumping the ring resonator off resonance and measuring the output power at the through port:

$$P_{\text{out,off}} = P_{\text{pump}}\alpha_{\text{wg}}\alpha_{c2} = P_{\text{in}}\alpha_{c1}\alpha_{\text{wg}}\alpha_{c2}. \quad \text{(S2)}$$

After shifting the laser into resonance and initiating the comb, the spectrum can again be measured at the through port giving us the on-resonance output powers:

$$P_{\text{out,on}} = P_{\text{out,pump}} + \sum P_{\text{out,comblines}} \quad \text{(S3)}$$
$$= P_{\text{out,pump}} + \alpha_{\text{wg}}\alpha_{c2} \sum P_{\text{comblines}}, \quad \text{(S4)}$$

where $P_{\text{out,pump}}$ is the power left in the pump line at the output. From equations S1, S2, and S4, we can now extract the conversion efficiency:

$$\eta_{\text{comb}} = \frac{P_{\text{out,on}} - P_{\text{out,pump}}}{P_{\text{in}}\alpha_{c1}\alpha_{\text{wg}}\alpha_{c2}} \quad \text{(S5)}$$
$$= \frac{P_{\text{out,on}} - P_{\text{out,pump}}}{P_{\text{out,off}}}. \quad \text{(S6)}$$

Using equation S6 we can thus measure the conversion efficiency of our comb by only looking at the through-port and taking optical spectra in on- and off-resonant pumping situations. A similar calculation can be done for the drop port, assuming the chip-to-fiber coupling losses are identical for the two ports giving effective conversion efficiencies for both output ports.

Figure S3b) shows the values for the combs that were used in our experiments. The two devices had slightly different gap distances between the ring and the through ports (500 nm vs 300 nm) while having the same 500 nm gap to the drop port leading to the difference in the ratio between the through and drop port conversion efficiencies.

**Fig. S2.** a) Schematic of the equipment used for the recirculating loop. b) Schematic of the equipment in the receiver.

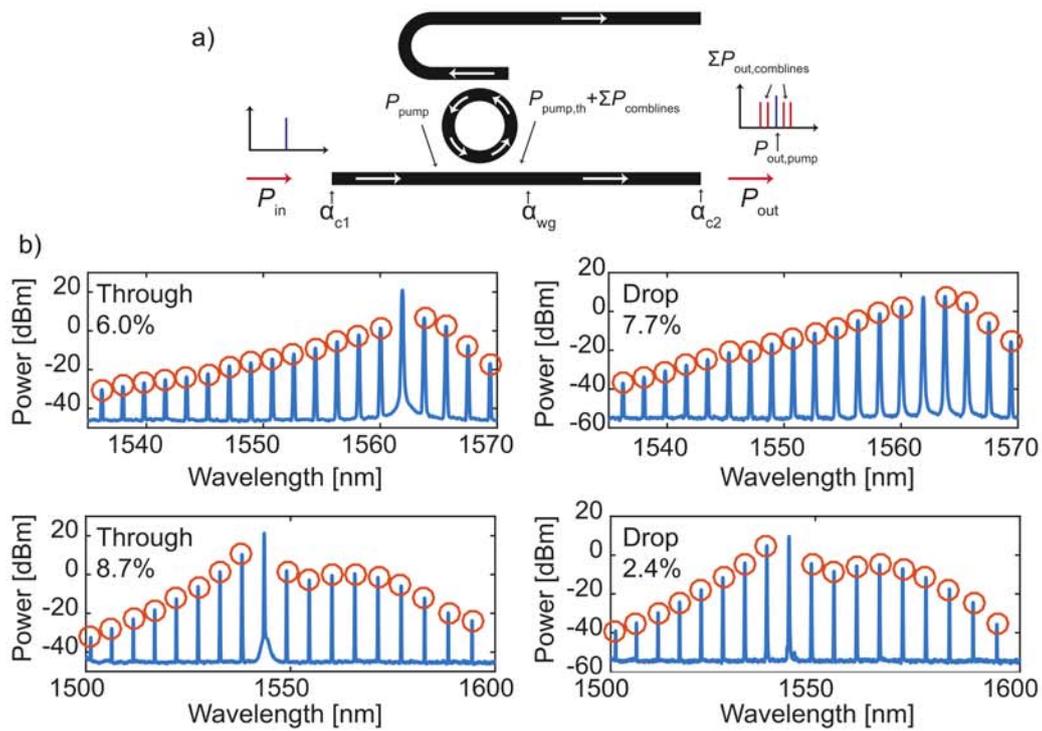

**Fig. S3.** a) Sketch of a microresonator displaying the measurable power locations ($P_{\text{in}}$ and $P_{\text{out}}$), the internal locations where the power levels need to be calculated as well as the loss elements. b) Recorded spectra from the through and the drop ports of the two used microresonators. The displayed conversion efficiencies were calculated by measuring the powers in all the lines except for the pump line and comparing it to the throughput power in the off-resonant pumping case. The powers were measured using a grating-based optical spectrum analyzer with 0.1 nm resolution. In both devices, the total conversion efficiency (adding the through and the drop port) exceeded 10 %.